\documentclass[fleqn,usenatbib]{mnras}

\usepackage{newtxtext,newtxmath}

\usepackage[T1]{fontenc}

\DeclareRobustCommand{\VAN}[3]{#2}
\let\VANthebibliography\thebibliography
\def\thebibliography{\DeclareRobustCommand{\VAN}[3]{##3}\VANthebibliography}

\usepackage{graphicx}	
\usepackage{amsmath}	

\usepackage{booktabs}
\usepackage{multirow}

\usepackage{cleveref}
\crefname{table}{Table}{Tables}
\crefname{figure}{Figure}{Figures}

\title[Massive White Dwarfs]{The  Cooling  of Massive White Dwarfs from \textit{Gaia} EDR3}

\author[Fleury, Caiazzo \& Heyl]{
Leesa Fleury\thanks{email: lfleury@phas.ubc.ca}$^1$,
Ilaria Caiazzo\thanks{email: ilariac@caltech.edu; Sherman Fairchild Fellow}$^2$,
Jeremy Heyl\thanks{email: heyl@phas.ubc.ca}$^1$
\\
$^{1}$Department of Physics and Astronomy, University of British Columbia, Vancouver, BC V6T 1Z1, Canada\\
$^{2}$TAPIR, Walter Burke Institute for Theoretical Physics, Mail Code 350-17, Caltech, Pasadena, CA 91125, USA
}

\date{Accepted XXX. Received YYY; in original form ZZZ}

\pubyear{2022}

\begin{document}
\label{firstpage}
\pagerange{\pageref{firstpage}--\pageref{lastpage}}
\maketitle

\begin{abstract}
We determine the distribution of cooling ages of massive \textit{Gaia} EDR3 white dwarfs identified with over 90\% probability within 200~pc and with mass in the range $0.95 - 1.25 ~M_\odot$. 
Using three sets of publicly available models, we consider sub-samples of these white dwarfs sorted into three equally spaced mass bins. 
Under the assumption of a constant white dwarf formation rate, we find an excess of white dwarfs both along the Q branch and below it, corresponding respectively to stars that are in the process of freezing and those that are completely frozen.  
We compare the cooling age distributions for each of these bins to the recently determined time-varying star formation rate of \textit{Gaia} DR2 main sequence stars. 
For white dwarfs in the two lightest mass bins, spanning the mass range $0.95-1.15~M_\odot$, we find that the cumulative cooling age distribution is statistically consistent with the expectation from the star formation rate. 
For white dwarfs in the heaviest mass bin, $1.15-1.25~M_\odot$, we find that their cumulative distribution is inconsistent with the star formation rate for all of the models considered; instead, we find that their cooling age distribution is well fitted by a linear combination of the distribution expected for single stellar evolution products and the distribution expected for double white dwarf merger products when approximately $40-50\%$ of the $1.15-1.25~M_\odot$ white dwarfs that formed over the past 4 Gyr are produced through double white dwarf mergers.
\end{abstract}




\section{Introduction}

The \textit{Gaia} mission \citep{2016A&A...595A...1G} has revealed the largest number of white dwarfs (WDs) ever observed in our galaxy.
Before \textit{Gaia}, less than 40,000 confirmed white dwarfs were known, mostly from the spectroscopic Sloan Digital Sky Survey \citep{2000AJ....120.1579Y,2006ApJS..167...40E}. This sample was largely incomplete and concentrated in the northern hemisphere. Thanks to \textit{Gaia}'s precise parallaxes and photometric measurements, \citet{2019MNRAS.482.4570G} were able to identify $\sim$260,000 high-confidence white dwarf candidates simply from their position in the HR diagram. The latest data release of \textit{Gaia}, \textit{Gaia} EDR3 \citep{2021A&A...649A...1G}, provided 20 to 30 percent better parallax measurements on average, as well as twice as accurate proper motions and better measurements of colour, especially for the bluest objects. These improvements allowed  \citet{GentileFusillo2021} to update the catalogue, which now includes $\sim$359,000 high-confidence white dwarfs.

The wealth of new data provided by \textit{Gaia} DR2 has challenged our understanding of white dwarf cooling, a process that was thought to be well understood.
The \textit{Gaia} DR2 colour-magnitude diagram (CMD) unveiled the presence of the Q branch, a transversal sequence of white dwarfs with mass above $\sim 1~M_\odot$ that was not aligned with theoretical cooling sequences \citep{2018A&A...616A..10G}, which \citet{Tremblay2019} identified as observational evidence of a white dwarf cooling delay due to core crystallization. 
While a pile-up of white dwarfs in the region of the Q branch is expected in standard crystallization models due to the cooling delay resulting from the release of latent heat in the phase transition from a liquid to solid state \citep{Tremblay2019}, \citet{Cheng2019} presented evidence of an additional anomolous cooling delay on the Q branch in \textit{Gaia} DR2 data experienced by massive ($1.08 - 1.23 M_{\odot}$) WDs that could not be explained by conventional models of core crystallization. 
This has resulted in much research into the details of WD crystallization, particularly $^{22}$Ne sedimentation, in an attempt to explain this cooling anomaly \citep{Bauer2020,Caplan2020,Blouin2020,Blouin2021,Camisassa2021}.

Another significant finding arising from \textit{Gaia} DR2 data was the discovery of a star formation burst in the galactic disc 2-3 Gyr ago based on observations of main sequence stars \citep{Mor2019}. In addition to presenting evidence of such an event, \citet{Mor2019} mathematically characterized the time dependent star formation rate (SFR), modelling it as a bounded exponential function plus a Gaussian component that accounts for the star formation burst.
\citet{Isern2019} noted the significance of this star formation history for the white dwarf luminosity function, reconstructing the SFR for 0.9 - 1.1 $M_\odot$ \textit{Gaia} DR2 white dwarfs within 100 pc of the Sun, using the luminosity function of \citet{Tremblay2019} and the BaSTI cooling models for DA white dwarfs and qualitatively comparing this to the SFR of \citet{Mor2019}.

The catalogue of \textit{Gaia} EDR3 WDs recently produced by \citet{GentileFusillo2021} provides us with a powerful tool to re-investigate the cooling of massive white dwarfs in the solar neighbourhood with improved statistical power.
In this paper, we determine the distribution of cooling ages for massive high-probability white dwarf candidates recently identified by the \citet{GentileFusillo2021} catalogue that are within 200~pc of the Sun. 
Using the publicly available \texttt{WD\_models} code \citep{sihaocheng}, we use different sets of white dwarf models, including both the carbon/oxygen (C/O) core cooling models of \citet{Bedard2020} and the oxygen/neon (O/Ne) core cooling models of \citet{Camisassa2019}, to determine the masses and cooling ages of each white dwarf candidate from the \textit{Gaia} EDR3 photometric observations.
For each set of models, we construct the distribution of cooling ages for objects with mass in the range $0.95-1.25~M_\odot$ and binned into subdivisions by mass within this range.
We quantitatively asses the statistical significance of the similarity of the cooling rate distribution of massive \textit{Gaia} EDR3 white dwarfs to the star formation rate of main sequence stars, as characterized by \citet{Mor2019}.
We also illustrate how the presence of a significant fraction of mergers among the most massive WDs, with single progenitors created at the \citet{Mor2019} rate of star formation, can explain the nearly uniform distribution that we find for their cooling ages.

\section{Methods}
\subsection{Models}

Throughout this work, we compare multiple sets of white dwarf cooling models to \textit{Gaia} EDR3 white dwarf data. For our analysis, we use the publicly available \texttt{WD\_models} package provided by \citet{sihaocheng}\footnote{The \texttt{WD\_models} package is publicly available at \url{https://github.com/SihaoCheng/WD_models}}. 
We consider the cooling models of the Montreal group \citep{Bedard2020} and the La Plata group \citep{Camisassa2019,Camisassa2017,Renedo2010}.
For each set of cooling models, we consider both H-atmosphere and He-atmosphere models. 
We use the evolutionary models in conjunction with the publicly available Montreal group synthetic colours\footnote{The Montreal group synthetic colours are publicly available at \url{http://www.astro.umontreal.ca/~bergeron/CoolingModels}.} 
for pure H and pure He atmosphere models with \textit{Gaia} EDR3 band-pass filters to determine the masses and cooling ages of each white dwarf candidate from the \textit{Gaia} EDR3 photometric observations.

For the Montreal group models \citep{Bedard2020}, we consider three cases of atmosphere composition: thick H envelope evolution models with a pure H atmosphere model (``thick H envelope models''), thin H envelope evolution models with a pure H atmosphere model (``thin H envelope models''), and thin H envelope evolution models with a pure He atmosphere model (``He envelope models''). 
We collectively refer to the set of thick H envelope models plus He envelope models as the ``thick Montreal'' models, and analogously refer to the set of thin H envelope models plus He envelope models as the ``thin Montreal'' models.
In all of these cases, the Montreal models have a C/O core composed of a uniform mixture of C and O in equal parts by mass ($X_C = X_O = 0.5$). 
It should be noted that this core composition differs from typical model predictions for single stellar evolution \citep[\textit{e.g.}][]{Siess2006,Lauffer2018,Althaus2021}, which predict lower fractions of C for WDs with mass in the range that we consider. 
Furthermore, the fixed uniform mixture used for the cores in the Montreal models does not account for element diffusion, which separates the elements into stratified layers.

For the La Plata group set of models, we combine the higher mass O/Ne core models of \citet{Camisassa2019} with the lower mass C/O core models of \citet{Renedo2010} and \citet{Camisassa2017}. While the \citet{Camisassa2019} models are available for both DA and DB WDs, the \citet{Renedo2010} models are only available for DA WDs and the \citet{Camisassa2017} are only available for DB WDs. 
Thus, for the La Plata group H-atmosphere models, we used the DA WD models of \citet{Camisassa2019} for masses $\geq 1.1 \ M_{\odot}$ and the models of \citet{Renedo2010} with metallicity $Z = 0.01$ for masses $\leq 0.93 M_{\odot}$, linearly interpolating between these models for masses between 0.93 and 1.1 $M_{\odot}$. To make the full set of He-atmosphere La Plata group models, we used the DB WD models of \citet{Camisassa2019} for masses $\geq 1.1 \ M_{\odot}$ and the models of \citet{Camisassa2017} for masses $\leq 1 \ M_{\odot}$.
We collectively refer to all of these models as the ``La Plata'' models.

The composition of ultramassive white dwarfs (with mass above $1.05~M_\odot$) is not well known.
The single stellar evolution simulations of \citet{Siess2006,Siess2007} and \citet{Doherty2010} indicate that white dwarfs transition from having C/O cores to O/Ne for masses $\gtrsim 1.05~M_\odot$, and \citet{Doherty2015} find this transition to correspond to masses between $\sim 1.07 - 1.15~M_\odot$ for solar metallicity. However, \citet{Althaus2021} showed that ultramassive C/O-core white dwarfs with masses in excess of $1.25~M_\odot$ can readily be produced by models with either high rotation rates or low rates of mass loss on the asymptotic giant branch (AGB).
Massive white dwarfs can also be formed through the merger of two lighter white dwarfs, and this process could result in white dwarfs with different composition than those produced from the evolution of a single progenitor \citep{Althaus2021}.
While the evolutionary models of \citet{Schwab2021} for the merger remnant of two C/O white dwarfs indicate that merger products with final mass $\gtrsim 1.05~M_\odot$ have O/Ne cores, other work \citep[\textit{e.g.}][]{Yoon2007,Loren-Aguilar2009} has found scenarios in which ultramassive remnants with C/O cores may be produced from double white dwarf mergers.

For ultramassive C/O white dwarf models, \citet{Althaus2021} noted that the composition of those produced by single stellar evolution and those produced by double white dwarf mergers is expected to be different.
WDs produced by single stellar evolution are expected to be born with the most abundant elements already largely separated into distinct layers due to element diffusion during pre-WD evolution, while the merger of two WDs mixes the elements and thus causes the resulting WD to (re)start cooling with a more uniformly mixed composition. 
Even for single stellar evolution models, there is significant uncertainty in the expected ratio of carbon to oxygen in C/O-core white dwarfs due to factors such as the uncertainty in the $^{12}C(\alpha,\gamma)^{16}O$ rate and the treatment of convection in progenitor models \citep[\textit{e.g.}][]{Salaris2010,DeGeronimo2017,Wagstaff2020}, in addition to the choice of rotation and mass loss on the AGB \citep{Althaus2021}.

We adopt an agnostic approach to the WD core composition by considering models with a variety of compositions and performing the same analysis for each set of models to determine the best-fitting model for each mass bin.
The models considered in this work are particularly relevant for comparison with the work of \citet{Cheng2019}, that found an anomalous cooling delay on the Q branch for massive \textit{Gaia} DR2 WDs.
The \citet{Bedard2020} models are the most recent version of the older \citet{Fontaine2001} models used by \citet{Cheng2019}, who used the thick H models of \citet{Fontaine2001} for WD masses $\leq 1.05~M_\odot$ and the models of \citet{Camisassa2019} for WD masses $\geq 1.1~M_\odot$.

\subsection{Data}

We used the main catalogue of \citet{GentileFusillo2021}, which consists of \textit{Gaia} EDR3 white dwarfs with reliable parallax measurements. From this catalogue, we selected sources within a distance of 200 pc and for which the probability of being a white dwarf exceeded 90\%, as determined by the \texttt{Pwd} parameter of the catalogue \citep[for details concerning how this probability was determined, see][]{GentileFusillo2021}. 

For each source in the catalgoue, \citet{GentileFusillo2021} provide the three parameters \texttt{chisq\_H}, \texttt{chisq\_He}, and \texttt{chisq\_mixed}, indicating the goodness-of-fit of models with different atmosphere compositions: pure H, pure He, and mixed H/He atmospheres, respectively. In our work, we classified each WD as having an atmosphere composition corresponding to the atmosphere model with the smallest (non-empty) chi-squared value. WDs for which all three chi-squared parameters were empty were classified as having an ``unknown'' atmosphere.
For a given set of models, this classification was used to determine whether the H-atmosphere or He-atmosphere models were used to determine the mass and cooling age of each source from the observed photometry. He-atmosphere models were used for both the pure He and mixed H/He atmosphere WDs. Following the recommendations of \citet{GentileFusillo2021}, we used H-atmosphere models for WDs with unknown atmosphere composition.

\begin{figure}
	\includegraphics[width=\columnwidth]{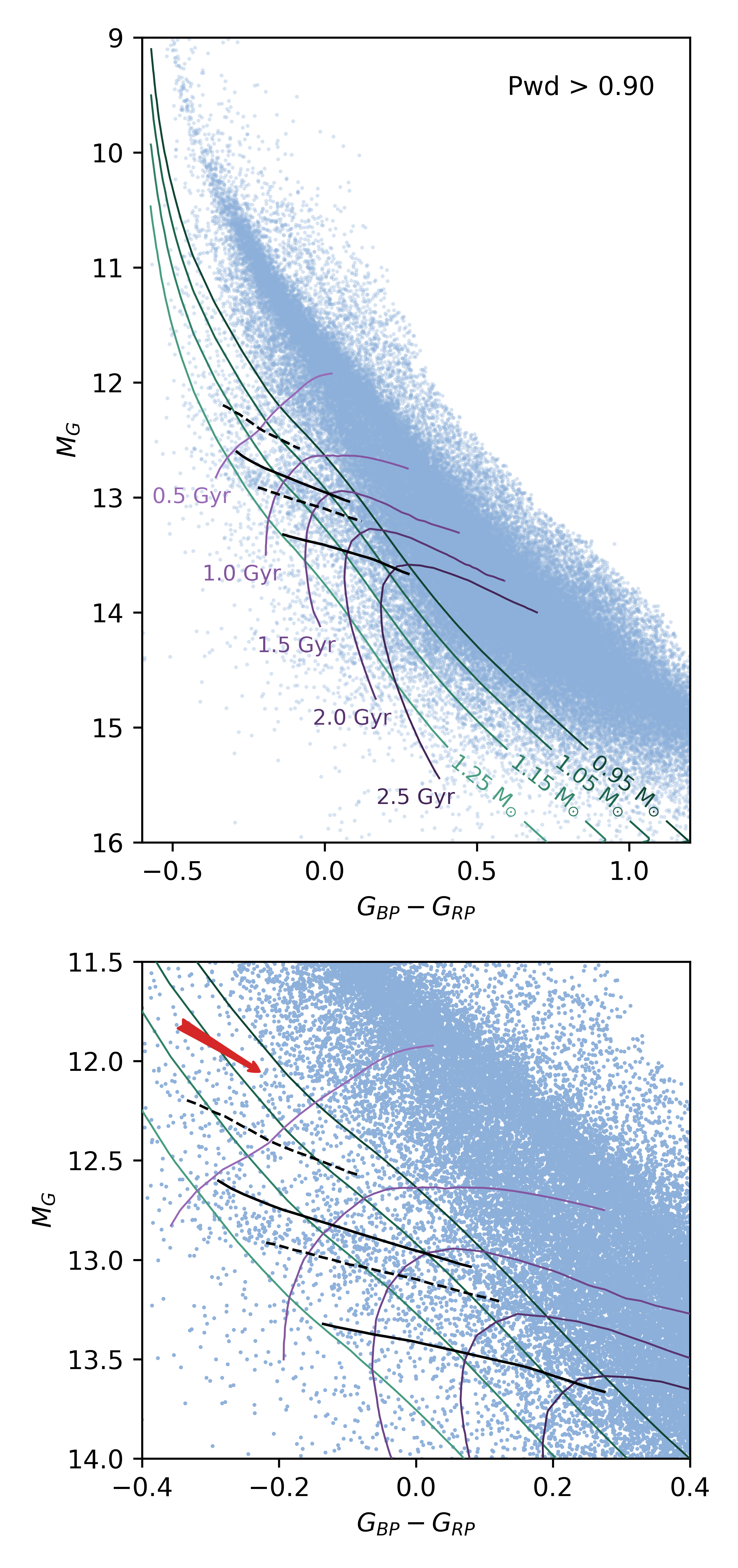}
    \caption{Upper: the \textit{Gaia} CMD of WD candidates within 200~pc from \citet{GentileFusillo2021}. Lower: focus on the high-mass white dwarfs. The effect of interstellar extinction with $E(B-V)=0.1$ is indicated by the red arrow. Superimposed on the observations are contours of equal mass, from 0.95 to 1.25 solar masses (right to left), and contours of equal cooling age, from 0.5 to 2.5~Gyr (top to bottom), according to the thin H atmosphere white dwarf cooling models of \citet{Bedard2020}. The four black lines show significant points in core crystallization, as identified by \citet{Bauer2020}. The dashed black lines indicate the region of O/Ne crystallisation, showing the points at which 20\% (top) and 80\% (bottom) of the O/Ne core is frozen. Analogously, the solid black lines show the points at which 20\% (top) and 80\% (bottom) of the C/O core is frozen.}
    \label{fig:wd_cmd}
\end{figure}

\Cref{fig:wd_cmd} depicts our observational sample of white dwarfs identified by \citet{GentileFusillo2021} within 200~pc.  Superimposed on the diagram are contours of constant mass and cooling age calculated using the thin H envelope models of \citet{Bedard2020}.
The other models considered in our work show similar trends. Furthermore, the thresholds and completion of core crystallisation for carbon/oxygen and oxygen/neon white dwarfs as identified by \citet{Bauer2020} are shown by solid and dashed black lines.  These contours are similar to those identified by \citet{Tremblay2019} for carbon/oxygen white dwarfs and to those by \citet{Camisassa2019} for oxygen/neon white dwarfs.  For the masses that we consider, the crystallisation occurs between the absolute G-band magnitude of 12 and 14. 
The magnitude and colour have been de-reddened using the mean $A_V$ values given in the catalogue and the prescription given by \citet{GentileFusillo2021} to convert this extinction in the Johnson V band to extinctions in \textit{Gaia} bands. To illustrate the shift in the CMD caused by reddening, we show the \textit{Gaia} extinction vector for a colour excess of $E(B-V)=0.1$ (the median extinction for stars within 200~pc from the \textit{Gaia} EDR3) as a red arrow in the lower panel of \cref{fig:wd_cmd}.
De-reddening undoes the effect illustrated by this arrow, shifting the WDs back in the opposite direction.
As the extinction vector is approximately parallel to the phase transition lines for the range of white dwarf masses that we will consider, the location of a white dwarf relative to the phase transition lines is approximately independent of extinction. While reddening will make a white dwarf appear slightly less massive by a few hundredths of a solar mass for $E(B-V)=0.1$, since we will be studying white dwarfs binned into samples one tenth of a solar mass wide, our conclusion should not be very sensitive to the exact treatment of reddening.

\begin{figure}
    \centering
    \includegraphics[width=\columnwidth]{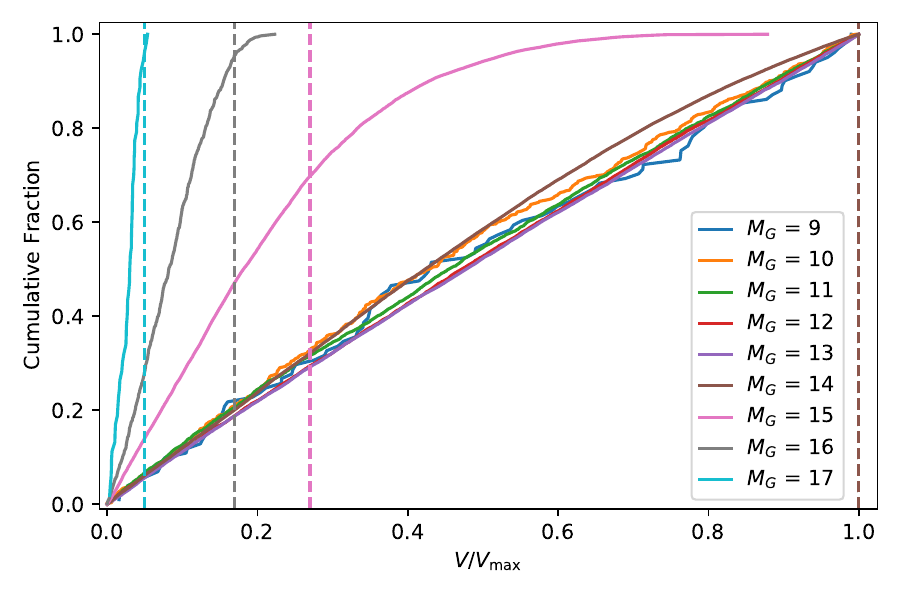}
    \caption{Completeness limits of white dwarf samples as a function of absolute magnitude. The vertical dashed lines indicate the volume limits of the complete samples that we consider as a function of absolute magnitude.}
    \label{fig:wd_complete}
\end{figure}

A key concern in determining the distribution of white dwarf cooling ages is to determine the volume sampled as a function of the luminosity of the white dwarf \citep{2021arXiv210607653R}.
Although we could estimate this using the magnitude limits of the \citet{GentileFusillo2021} catalogue, we would still not have a reliable estimate of the completeness rate at the faint end of the catalogue; consequently, we will study the completeness of the catalogue as a function of absolute G-band magnitude and distance using a variant of the \citet{1968ApJ...151..393S} estimator.  If the population that we are sampling is uniform surrounding the Sun (which is a good approximation within 200~pc), the number of white dwarfs that we detect should increase linearly with the volume sampled. \Cref{fig:wd_complete} depicts the cumulative number of white dwarfs detected as a function of the volume sampled (normalised by the total volume $V_\textrm{max}$, which is the volume of a sphere with a 200 pc radius).  White dwarfs with absolute magnitudes of 14 and brighter are detected with high completeness all the way out to 200~pc.  For inherently fainter white dwarfs, the samples become incomplete closer to the Sun. The incompleteness manifests itself as the downward curvature of the cumulative distribution from the linear relation.
We choose the limiting volume to be where this curvature develops.  In particular we choose the limiting volume to be where the slope of the cumulative distribution deviates by more than five percent relative to its value for small volumes.  Consequently, within these limiting volumes, the samples are nearly complete.  We only perform this adjustment for samples that we estimate to be less than 85\% complete within 200~pc, that is, for $M_G\geq 15$.
The three leftmost vertical lines from left to right show the limiting volume that we choose for $M_G=17, 16$ and $15$. 
For $M_G \leq 14$, we do not reduce the volume relative to the total volume of the 200~pc sample, and we illustrate this choice as a limiting volume equal to $V_\textrm{max}$ shown by the rightmost vertical line in \cref{fig:wd_complete}.
We restrict our sample to objects within their magnitude dependent completeness-limiting volume, $V_\mathrm{lim}(M_G)$.
The reduced sampling volume of the faint white dwarfs is then corrected by assigning a weight of $V_{\mathrm{max}} \ / \ V_{\mathrm{lim}}(M_G)$ to objects for which $V_{\mathrm{lim}}(M_G) < V_{\mathrm{max}}$ and a weight of 1 to objects for which $V_{\mathrm{lim}}(M_G) \geq V_{\mathrm{max}}$.

\section{Results}

\begin{figure*}
    \centering
    \includegraphics[width=\textwidth]{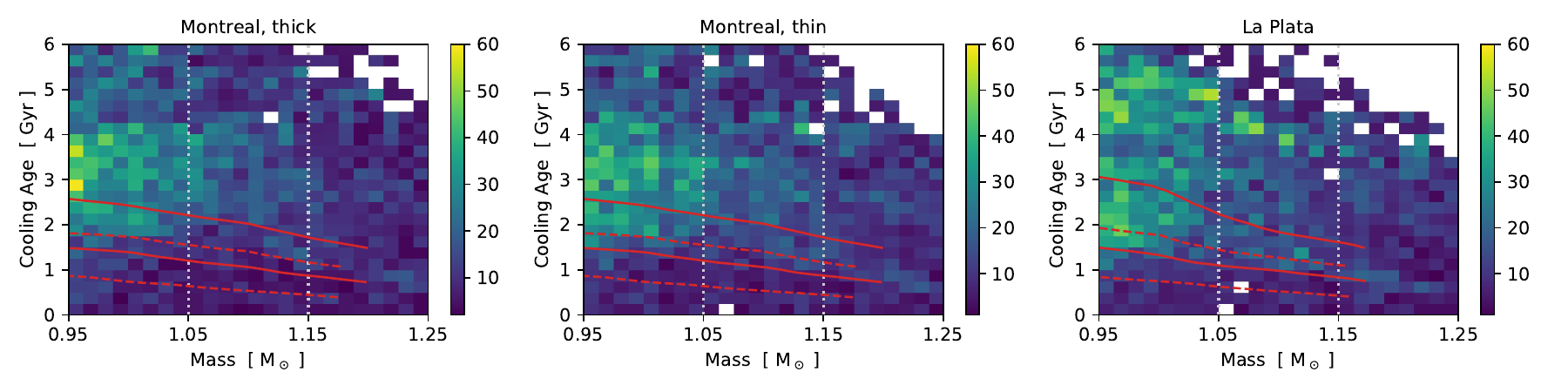}
    \caption{Joint distribution of mass and cooling age for massive \textit{Gaia} EDR3 WDs according to different sets of models, weighted to correct for reduced sampling volume. For each subplot, the set of models used to calculate the cooling ages and masses from \textit{Gaia} EDR3 photometry is indicated at the top of the subplot. From left to right, these are the thick Montreal models, the thin Montreal models, and the La Plata models (see main text for details). The fill colour of each bin indicates the corrected count found in that bin, with corresponding value given by the colour bar and white denoting that no objects were found in that bin. The red lines indicate the C/O (solid) and O/Ne (dashed) freezing lines of \citet{Bauer2020}, shown only for reference. From bottom to top, these lines indicate 20\% O/Ne frozen, 20\% C/O frozen, 80\% O/Ne frozen, and 80\% C/O frozen. The vertical dotted lines indicate the boundaries between the three mass bins considered in this work.}
    \label{fig:2dhist}
\end{figure*}

In \cref{fig:2dhist} we show the joint distribution of masses and cooling ages of \textit{Gaia} EDR3 WDs according to different sets of models. To generate these distributions, the WDs were weighted to correct for the reduced sampling volume required to make the sample complete.
For each set of models, the mass and cooling age of each object in our reduced sample was determined by linearly interpolating the models, using either the H-atmosphere or He-atmosphere models depending on the atmosphere classification of the source (as determined by the \texttt{chisq\_H}, \texttt{chisq\_He}, and \texttt{chisq\_mixed} parameters of the \citet{GentileFusillo2021} catalogue). 
Later in this work we sort the WDs into three mass bins of width 0.1 $M_\odot$, centered on the values 1.0, 1.1, and 1.2 $M_\odot$. The edges of these mass bins are indicated in \cref{fig:2dhist} by light grey dotted vertical lines. 
For illustrative purposes and to help orient the reader, we also show the core crystallization lines of \citet{Bauer2020} as red lines in \cref{fig:2dhist}. The solid lines indicate C/O crystallization thresholds of 20\% (bottom, solid) and 80\% (top, solid) frozen, while the dashed lines similarly indicate O/Ne crystallization thresholds of 20\% (bottom, dashed) and 80\% (top, dashed) frozen. \citet{Bauer2020} calculated these crystallization thresholds using models with pure He atmospheres and \textit{Gaia} DR2 filters. 
For consistency with the \citet{Bauer2020} models used to calculate these threshold, we used He-atmosphere models to convert those (\textit{Gaia} DR2) colour-magnitude crystallization thresholds to curves in mass and cooling age. The models used for this conversion consisted of the He-envelope evolution models of each of the three sets of models considered in this work along with the synthetic colours for \textit{Gaia} DR2 filters, rather than those for the EDR3 filters used to determine the masses and cooling ages of objects in our observational sample.
By comparison with the crystallization lines in \cref{fig:wd_cmd}, we can identify where the Q branch region of the colour-magnitude diagram has been mapped to in \cref{fig:2dhist}, as the bulk of the Q branch region occurs between the lines of 20\% C/O frozen and 80\% O/Ne frozen.

Several trends are immediately apparent from \cref{fig:2dhist}, regardless of which set of models is considered. The decreasing density from left to right across each plot indicates that there are more low-mass WDs than high-mass WDs.
Equally striking is that, as the cooling age increases (moving from bottom to top in each plot), the density of WDs gradually increases until it reaches a peak well after the age at which 80\% of a C/O core is frozen. 
This corresponds to an excess of stars on the Q branch and below it on the colour-magnitude diagram. This effect is much more pronounced for the lower mass WDs, with masses $\sim 0.95 - 1.15~M_\odot$.
Since WDs with mass $\gtrsim 1.15~M_\odot$ are much less numerous than WDs on the lower end of the mass range shown in \cref{fig:2dhist}, the scale of the plots makes it difficult to see the trend with cooling age for these most massive WDs; however, our analysis later in this work will reveal that their cooling ages are much closer to uniformly distributed.

If the models are correct and the WD birthrate were constant, then the distribution of cooling ages at a given mass would be expected to be uniform. The non-uniform density seen in \cref{fig:2dhist} (for $0.95-1.15~M_\odot$ WDs) indicates that either the WD birthrate is not constant or that these WDs are cooling more slowly than predicted by the models in the region where the density increases. 
In the latter scenario, the observed pile-up of WDs would imply a cooling delay that spans both the Q branch region and well beyond it, with the apparent delay in fact being much more prominent for WDs that have already finished their Q branch stage of evolution. 
The peak of this WD pile-up actually roughly coincides with the burst of star formation found by \citet{Mor2019} for \textit{Gaia} DR2 main sequence stars, with the general trend of the density gradually increasing with cooling age to a peak for the first $\sim 3~\mathrm{Gyr}$ seeming to follow the corresponding time-varying star formation rate found in that work. This strongly suggests that the non-uniform distribution of the WD cooling ages seen in \cref{fig:2dhist} is simply the imprint of the time-varying star formation history of their main sequence progenitors rather than an indication of a cooling delay relative to the standard WD cooling models.
Progenitor stars that produce massive WDs in the mass range that we consider ($0.95-1.25~M_\odot$) evolve through their pre-WD stages of evolution on timescales that are negligible compared to the cooling ages we consider, with pre-WD lifetimes of less than $100~\mathrm{Myr}$ \citep{2016ApJ...823..102C,pleiades}. Thus, if the massive WDs that we consider are born from single progenitors, then the distribution of their cooling ages should follow the star formation rate of their main sequence progenitors. 

\begin{figure*}
    \centering
    \includegraphics[width=\textwidth]{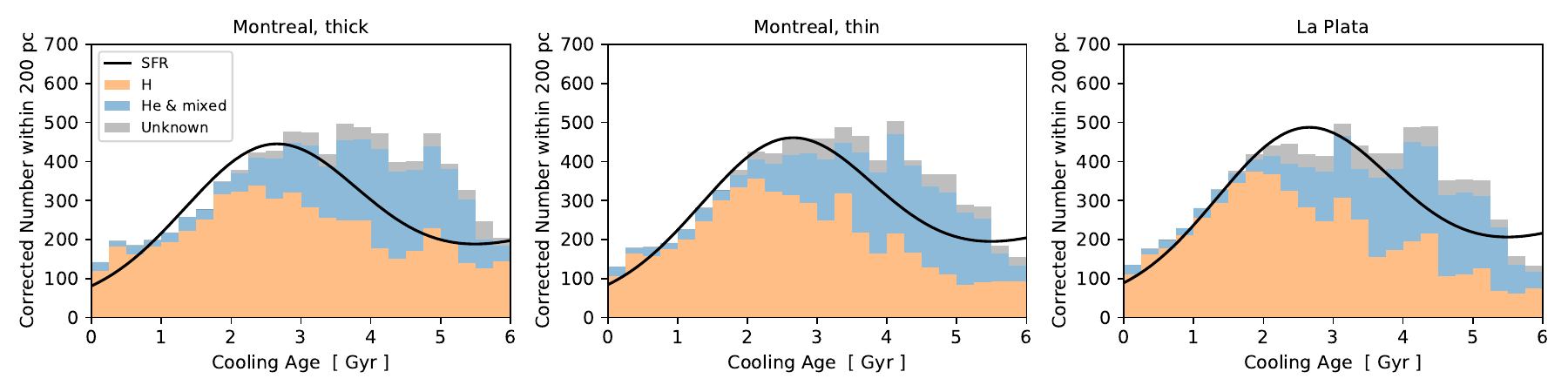}
    \caption{Distribution of cooling ages (in Gyr) of 0.95 - 1.25 $M_{\odot}$ WDs within 200 pc of the Sun, weighted to correct for reduced sampling volume.
    This sample was sorted into subgroups based on atmospheric composition, as determined by the fits of \citet{GentileFusillo2021}. WDs with atmospheric composition classified as either pure He or mixed H/He were combined into one group (shown in blue), and He-atmosphere models were used to determine their ages and masses. H-atmosphere models were used to determine the ages and masses of WDs with atmospheric composition classified as either pure H or unknown, but the H-atmosphere WDs (shown in orange) and the unknown-atmosphere WDs (shown in grey) were kept as separate groups. The (corrected) counts for each of these groups are stacked such that the top of the stacked distribution indicates the total (corrected) count from combining all three groups. The black line indicates the star formation rate of main sequence stars determined by \citet{Mor2019}, normalized to the number of WDs with cooling age $\leq$ 3 Gyr.}
    \label{fig:1dhist}
\end{figure*}

The distribution of cooling ages for WDs in the mass range 0.95 - 1.25 $M_\odot$, according to different sets of models and weighted to correct for the completeness-limiting reduced sampling volume, are shown in \cref{fig:1dhist}. 
These plots are the result of marginalizing the joint distributions of \cref{fig:2dhist} over WD mass for the given mass range.
The different colours indicate the contribution of WDs with different atmosphere classifications to the total distribution.
Nearly all of the youngest WDs, with cooling age $\lesssim 2~\mathrm{Gyr}$, are classified as having a H-atmosphere; while there are significant numbers of both H-atmosphere and He-atmosphere WDs with older cooling ages. WDs with unknown atmosphere composition only make up a very small fraction of the total WDs at any cooling age, so our treatment of these WDs should not significantly impact our results.

In each panel of \cref{fig:1dhist}, we also show the star formation rate determined by \citet{Mor2019} for \textit{Gaia} DR2 main sequence stars, which we have normalized to the number of WDs with a cooling age of $\leq$ 3~Gyr. Up to about 3 Gyr, the distribution of massive \textit{Gaia} EDR3 WDs tracks the shape of the SFR distribution, up to a small lag that should be expected due to the time required for a main sequence star to become a white dwarf. \citet{pleiades} have measured the lag between star formation and the formation of a one-solar-mass white dwarf to be less than 100~Myr from the Pleiades, so neglecting this lag is a reasonable approximation.
If the WD birthrate were constant, this distribution should be approximately uniform.
Instead we see an accumulation of massive WDs at cooling ages that coincide with the look-back time for the star formation burst found by \citet{Mor2019}.
Beyond about 3~Gyr, the WD cooling age distribution appears to track the SFR less well. This could indicate that the star formation burst is actually broader than found for the fiducial case of \citet{Mor2019}. 
The discrepancy for old WDs could alternately be due to the poorer statistics of the oldest WDs or an imperfect correction for the reduced sampling volume of these stars, which are typically fainter and less complete than the younger white dwarfs. The weighting only begins to affect the distributions for cooling ages older than about 3~Gyr, so we can reliably compare the cooling age distribution to the SFR up to at least 3 Gyr even if the latter situation is the issue.
We examine the divergence between the weighted and unweighted distribution more closely in our analysis of the cumulative cooling age distributions below.

\begin{figure*}
    \centering
    \includegraphics[width=\textwidth]{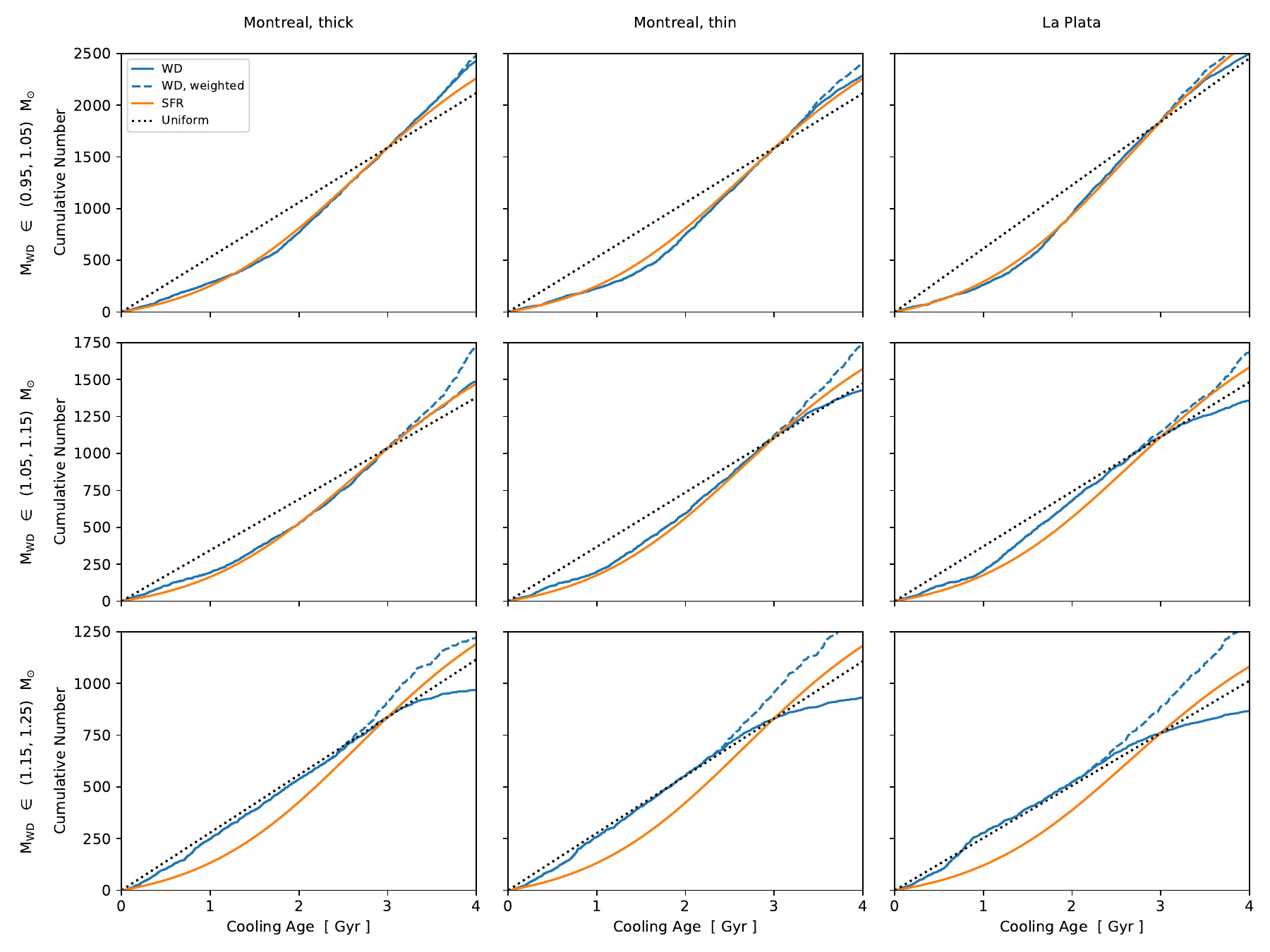}
    \caption{Cumulative distributions of cooling ages (in Gyr) of massive WDs in different mass bins according to the different sets of cooling models considered in this work. Each column corresponds to a particular set of models: thick Montreal (left), thin Montreal (centre), and La Plata (right). Each row corresponds to a particular WD mass bin: $0.95 - 1.05~M_\odot$ (top), $1.05 - 1.15~M_\odot$ (middle), and $1.15 - 1.25~M_\odot$ (bottom). For each subplot, we show both the unweighted (solid blue line) and weighted (dashed blue line) WD distributions. For comparison, we also show the cumulative distribution for the star formation rate of \citet{Mor2019} (solid orange line) and a uniform distribution (dotted black line), both normalized to the (unweighted) WD distribution at 3 Gyr.}
    \label{fig:cumdists_mbins}
\end{figure*}

To investigate the effect of white dwarf mass, $M_{\mathrm{WD}}$, on the distribution of cooling ages, we sort the massive WDs into three mass bins: 0.95 - 1.05 $M_{\odot}$, 1.05 - 1.15 $M_{\odot}$, and 1.15 - 1.25 $M_{\odot}$. The cumulative distributions for WDs in each of these mass bins, according to the three sets of models we consider, are shown in \cref{fig:cumdists_mbins}. We show the WD distribution both with and without weighting to correct for the reduced sampling volume for which the \textit{Gaia} EDR3 observations are complete. In each panel of \cref{fig:cumdists_mbins}, we show the cumulative distribution corresponding to the SFR of \citet{Mor2019}, normalized to the unweighted WD distribution at 3 Gyr. For reference, we also show a uniform distribution with the same normalization.
From the top and middle rows of this figure, we see that the distributions for WDs in the two lightest mass bins, $0.95-1.05~M_\odot$ and $1.05-1.15~M_\odot$, approximately follow the time-dependent SFR of main sequence stars \citep{Mor2019}. 
In stark contrast, the distribution of the heaviest WDs, in the mass bin $1.15-1.25~M_\odot$ (bottom row), does not seem to follow the SFR at all; instead, the cooling ages of the heaviest WDs are nearly uniformly distributed.

To statistically asses the similarity between the observed cumulative cooling age distributions and the cumulative SFR of \cite{Mor2019}, we perform a series of one-sample Kolmogorov–Smirnov (KS) tests. 
We test the null hypothesis that the empirical cumulative distribution of a particular sample of WD cooling ages is equal to the analytic cumulative distribution function of the SFR.
These cumulative distributions will be equivalent if the models are correct and the WD birthrate is proportional to the SFR of the main sequence progenitors, \textit{i.e.} if the probability density of producing a WD with a particular cooling age is proportional to the SFR at the equivalent look-back time.
Viewing the cooling age of a WD as a random variable drawn from an underlying probability distribution, which here is taken to be proportional to the SFR under the null hypothesis, the p-value from the one-sample KS tests quantifies the probably of randomly drawing a sample from the proposed underlying distribution for which the sample distribution differs from the underlying distribution by at least as much as what is seen for the observed empirical distribution.
A small p-value indicates that the observed WD cooling ages are unlikely to have been drawn from the proposed SFR distribution. Viewed as a goodness-of-fit test, a small p-value indicates a poor fit of the SFR to the empirical distribution of cooling ages.

\begin{table}
	\centering
	\caption{The p-values from one-sample KS tests comparing the (unweighted) cumulative distribution of cooling ages of WDs in different mass bins, according to different models, to the cumulative SFR of \citet{Mor2019} for samples restricted to different maximum cooling ages, $t_\mathrm{max}$ (in Gyr).}
	\label{tab:KS_results_massbins}
	\begin{tabular*}{\columnwidth}{lcrrr}
	\toprule
    Mass ($M_\odot$)    &  $t_\mathrm{max}$ (Gyr)  &  Montreal, thick  &  Montreal, thin  &  La Plata  \\
    \midrule
    $0.95 - 1.05$  &  2  &  0.0005  &  $<10^{-4}$  &  $<10^{-4}$  \\
                   &  3  &  0.0177  &  $<10^{-4}$  &  0.0266  \\
                   &  4  &  $<10^{-4}$  &  $<10^{-4}$  &  $<10^{-4}$  \\
    \addlinespace
    $1.05 - 1.15$  &  2  &  $<10^{-4}$  &  0.0103  &  0.0179  \\
                   &  3  &  0.0138  &  0.0208  &  $<10^{-4}$  \\
                   &  4  &  0.0707  &  $<10^{-4}$  &  $<10^{-4}$  \\
    \addlinespace
    $1.15 - 1.25$  &  2  &  $<10^{-4}$  &  $<10^{-4}$  &  $<10^{-4}$  \\
                   &  3  &  $<10^{-4}$  &  $<10^{-4}$  &  $<10^{-4}$  \\
                   &  4  &  $<10^{-4}$  &  $<10^{-4}$  &  $<10^{-4}$  \\
    \midrule
    $0.95 - 1.25$  &  2  &  $<10^{-4}$  &  $<10^{-4}$  &  0.0033  \\
                   &  3  &  $<10^{-4}$  &  0.0008  &  $<10^{-4}$  \\
                   &  4  &  $<10^{-4}$  &  $<10^{-4}$  &  $<10^{-4}$  \\
    \bottomrule
	\end{tabular*}
\end{table}

We performed separate KS tests for each mass bin and each set of models, as well as for the full mass range $0.95-1.25~M_\odot$ encompassing all bins.
As can be seen in \cref{fig:cumdists_mbins}, the weighted distribution starts to deviate from the unweighted distribution at cooling ages $\gtrsim$ 3~Gyr for the first two mass bins, and slightly earlier for the most massive bin. The KS tests do not account for the weighting of the WDs, so we can only use these tests to draw statistically meaningful conclusions for distributions of WDs young enough that the weighted distribution does not yet differ significantly from the unweighted distribution. For each sample we consider in a KS test, the sample includes only WDs with cooling age $\leq t_\mathrm{max}$. For each combination of model set and mass range, we perform KS tests for three choices of maximum cooling age: 2~Gyr, 3~Gyr, and 4~Gyr. 
The p-values resulting from all of these tests are summarized in \cref{tab:KS_results_massbins}.
These different samples allow us to consider different options for balancing the improved statistical power of larger sample numbers with the error for the oldest WDs from not accounting for the reduced sampling volume. For the lightest two mass bins, we consider the sub-samples for which the WD cooling ages are $\leq 3~\mathrm{Gyr}$ to give the best balance between these two considerations.

For each of the two lightest mass bins, there are two sets of models that give results consistent with the SFR for the sample with $t_\mathrm{max} = 3~\mathrm{Gyr}$, though the optimal models are different for each bin.
The La Plata models give the best results for the $0.95 - 1.05~M_\odot$ mass bin, with a p-value of 0.0266 for $t_\mathrm{max} = 3~\mathrm{Gyr}$, while the thick Montreal models give the best results for the $1.05 - 1.15~M_\odot$ mass bin, with a p-value of 0.0138 for $t_\mathrm{max} = 3~\mathrm{Gyr}$ (and 0.0707 for $t_\mathrm{max} = 4~\mathrm{Gyr}$).
For WDs in the heaviest mass bin, $1.15 - 1.25~M_\odot$, none of the models give WD distributions consistent with the SFR (all p-values $< 10^{-4}$).
This result is in line with \cref{fig:cumdists_mbins}, which shows that these heaviest WDs have approximately uniformly distributed cooling ages, particularly for cooling ages above $\sim 1$~Gyr, regardless of which set of models is considered.
To assess this apparently uniform trend, we performed another series of one-sample KS tests for just WDs in this heaviest bin to compare their cooling age distribution to a uniform distribution. 
For distributions over the cooling age range $1-2.5~\mathrm{Gyr}$, all of the models give distributions that are highly consistent with a uniform distribution, with p-values for KS tests comparing these distributions to a uniform distribution of 0.9803, 0.6321, and 0.2523 for the thick Montreal, thin Montreal, and La Plata models, respectively.
The distributions for the first 1~Gyr (\textit{i.e.} restricted to the cooling age range $0-1~\mathrm{Gyr}$) are in tension with a uniform distribution, having p-values of 0.0014, 0.0005, and $<10^{-4}$. This indicates that it could take $\sim 1~\mathrm{Gyr}$ for the uniform distribution to develop.

A possible explanation of the approximately uniform distribution of the most massive WDs is that a large fraction of these WDs formed due to mergers.
\citet{Temmink2020} have used binary population synthesis simulations to show that the assumption of single stellar evolution for white dwarfs that have undergone mergers typically leads to an underestimate of the white dwarf age, particularly for white dwarfs more massive than $0.9~M_\odot$ that are produced by double white dwarf mergers.
\citet{Temmink2020} further estimated that $30-50\%$ of white dwarfs with mass above $0.9~M_\odot$ are the product of binary mergers.
The earlier binary population synthesis simulations of \citet{Bogomazov2009} predicted even higher merger fractions for the most massive white dwarfs, with over 50\% of white dwarfs more massive than $1.1~M_\odot$ predicted to be the product of double white dwarf mergers.
These theoretical predictions of a high fraction of merger products among massive white dwarfs have been supported by observations.
Based on the kinematic, magnetic, and rotational properties of white dwarfs in the Montreal White Dwarf Database 100~pc sample, \citet{Kilic2021} found direct observational evidence that at least 32\% of the ultra-massive white dwarfs in their sample, with mass $>1.3~M_\odot$, were the product of binary mergers.
\citet{Cheng2020} have shown strong evidence of the presence of double-WD merger products among \textit{Gaia} DR2 high mass WDs and determined the cooling delay time distribution of double-WD merger products for multiple mass bins using population synthesis simulations. 

For the mass ranges we consider, if most WDs in the sample were formed from the evolution of a single progenitor, then the distribution of WD cooling ages would be expected to directly track the SFR of their main sequence progenitors, similar to what we see for the lightest two mass bins. However, WDs that are the result of a double-WD merger event will have a younger apparent cooling age than predicted by single WD evolution models.
If a large fraction of WDs in the sample are the result of double-WD merger events, then their cooling ages will be approximately distributed according to the convolution of the SFR of the single progenitors and the distribution of cooling delay times due to mergers.

\begin{figure}
    \centering
    \includegraphics[width=\columnwidth]{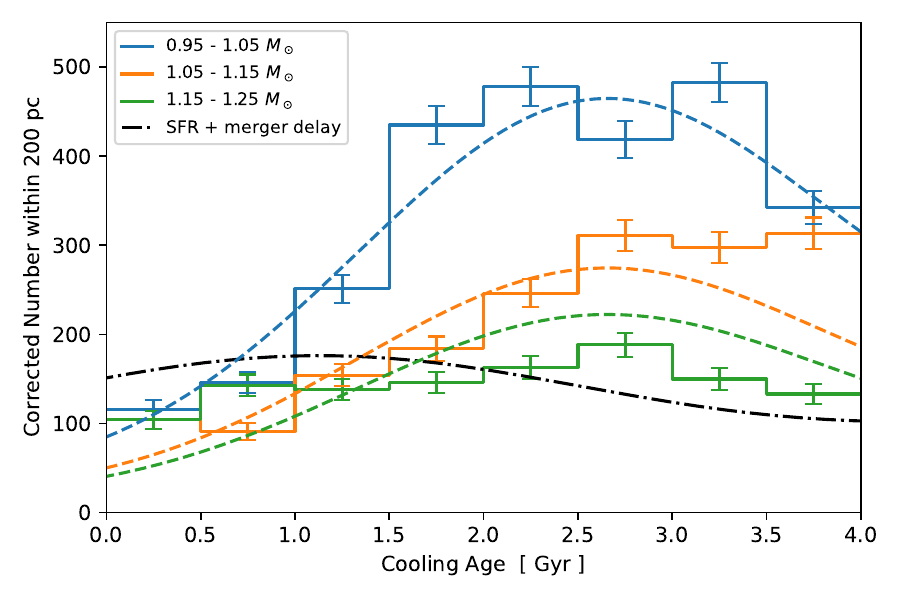}
    \caption{Distribution of white dwarf cooling ages by mass bin, weighted to correct for the reduced sampling volume of the complete sample. For the lightest two mass bins, we show the distributions made using the set of models that was most consistent with the star formation rate based on the results of KS tests. For the heaviest bin, we used the thick Montreal models. The dash-dot black curve indicates the distribution resulting from convolving the star formation rate of \citet{Mor2019} with the merger delay distribution of \citet{Cheng2020} appropriate for our heaviest mass bin; this curve has been normalized to the number of WDs in the heaviest bin with cooling age $\leq 3$~Gyr. The dashed lines indicate the star formation rate of \citet{Mor2019} analogously normalized to the number of WDs with cooling age $\leq 3~\mathrm{Gyr}$ in the mass bin with the corresponding colour. From top to bottom these are the lightest to heaviest mass bins.
    }
    \label{fig:mergers_cnts_all}
\end{figure}

To estimate the cooling age distribution if most of the WDs in our most massive bin are the result of double-WD mergers, we convolve the \citet{Cheng2020} distribution of cooling delay times for $1.14-1.24~M_\odot$ (approximately corresponding to our mass bin) with the \citet{Mor2019} SFR.
The result of this convolution is shown as the solid black curve in \cref{fig:mergers_cnts_all}, normalized to the number of WDs with apparent cooling age $\leq 3~\mathrm{Gyr}$ in our heaviest bin according to the thick Montreal models.
The dashed coloured curves in \cref{fig:mergers_cnts_all} show the SFR by itself, without convolution with the merger delay time distribution, with each curve normalized to the number of WDs with cooling age $\leq 3~\mathrm{Gyr}$ in the mass bin of the corresponding colour. Each normalization was determined using the most statistically consistent model for that mass bin according to the KS tests.
Note that normalizing the SFR to a different mass bin only affects the height of the curve, not the shape. 

The cooling age distributions for the observed \textit{Gaia} EDR3 WDs in each mass bin, according to the most consistent set of models for each bin as determined by the earlier KS tests, are shown in \cref{fig:mergers_cnts_all} as histograms with 0.5~Gyr bin width. These distributions were calculated using the same weighting as the previous figures in this work to correct for the reduced sampling volume of the complete samples.
The La Plata models were used for the lightest bin and the thick Montreal models for the two heaviest bins.
Comparison of the dashed green and solid black curves in \cref{fig:mergers_cnts_all} illustrates the effect on the cooling age distribution of accounting for the merger delay, shifting the peak associated with the star formation burst event to younger cooling ages and overall flattening the distribution, resulting in a distribution much closer to uniform and in better agreement with the observed distribution for WDs in the most massive bin.

We model the cooling age distribution of ultramassive white dwarfs as a linear combination of the distribution for white dwarfs produced by direct stellar evolution and the distribution for white dwarfs produced by the merger of two lighter white dwarfs.
The distribution for white dwarfs that evolved from a single progenitor is taken to be the SFR found by \citet{Mor2019} for main sequence stars, while the distribution for merger products is taken to be the convolution of the SFR of \citet{Mor2019} and the cooling delay distribution determined by \citet{Cheng2020}. 
We determine the empirical cooling age distribution for $1.15-1.25~M_\odot$ WDs according to each set of WD cooling models by constructing weighted histograms of 0.5 Gyr bin width for cooling ages spanning 0 to 4 Gyr, and we fit this model to each empirical distribution using non-linear least squares regression. 
The results of these fits are summarized in \cref{tab:merger_combo_fits}.
We also perform fits for the two limiting cases in which the ultramassive WDs are only the product of mergers and in which they are only the product of single stellar evolution, labelled respectively as ``Only Mergers'' and ``Only Singles'' in \cref{tab:merger_combo_fits}.

\begin{table}
    \centering
    \caption{
    Fitting results for cooling age distribution of $1.15-1.25~M_\odot$ WDs.
    The goodness-of-fit, scaled to the degrees of freedom ($\nu$), for each combination of origin model and WD cooling model is indicated by the reduced chi-squared value, $\chi^2_\nu$.
    We fit origin scenarios in which all of the WDs are the result of mergers (``Only Mergers'', $\nu=7$), all are the result of single stellar evolution (``Only Singles'', $\nu=7$), and some fraction are merger products (``Combination'', $\nu=6$).
    The fraction of merger products, $f_m$, forming today and averaged over 4 Gyr is given for each best-fit combination.
    }
    \begin{tabular*}{\columnwidth}{llccc}
    \toprule
    ~ & ~ & Montreal, thick & Montreal, thin & La Plata \\
    \midrule
    \multirow{3}{*}{$\chi^2_\nu$} & Combination & 1.4 & 4.9 & 11.6 \\
     & Only Mergers & 10.5 & 17.6 & 26.2 \\
     & Only Singles & 11.3 & 12.6 & 15.3 \\
     \midrule
     \multirow{2}{*}{$f_m$} & Today & 0.805 & 0.756 & 0.688 \\
     & Average & 0.511 & 0.439 & 0.358 \\
     \bottomrule
    \end{tabular*}
    \label{tab:merger_combo_fits}
\end{table}

The goodness-of-fit is assessed using the reduced chi-squared statistic, $\chi^2_\nu$, which allows us to compare the fits for the different distributions despite differences in the degrees of freedom, $\nu$. 
Fits for the limiting cases of only merger products and only single stellar evolution products each have $\nu=7$, while fits for the fiducial combination scenario have $\nu=6$. 
Even after accounting for the different degrees of freedom, the combination scenario gives a better fit than either of the limiting cases, regardless of which WD cooling model is used to determine the masses and cooling ages. 
Among the different WD cooling models, the thick Montreal models give the best fit for all of the cooling age distributions.
For the best-fitting combination of merger and direct formation products according to each set of WD cooling models, we determine the fraction of merger products, $f_m$, as a function of cooling age. 
In \cref{tab:merger_combo_fits}, we summarize $f_m$ by reporting the fraction of merger products among WDs forming today (at 0~Gyr cooling age) and the fraction of merger products among WDs that have formed over the past 4~Gyr.
The best-fitting cooling models, the thick Montreal models, give merger product fractions of $0.805 \pm 0.037$ for WDs forming today and $0.511 \pm 0.059$ for WDs that have formed over the last 4~Gyr, in line with the theoretical estimates of \citet{Bogomazov2009} and the observational results of \citet{Kilic2021}, though somewhat larger than the observational results of \citet{Cheng2020}.
From the $\chi^2_\nu$ values summarized in \cref{tab:merger_combo_fits}, we see that the case of the combination formation scenario with the thick Montreal cooling models fits the data well, with $\chi^2_\nu = 1.4$, but none of the other cases produces a good fit.

\begin{figure}
    \includegraphics[width=\columnwidth]{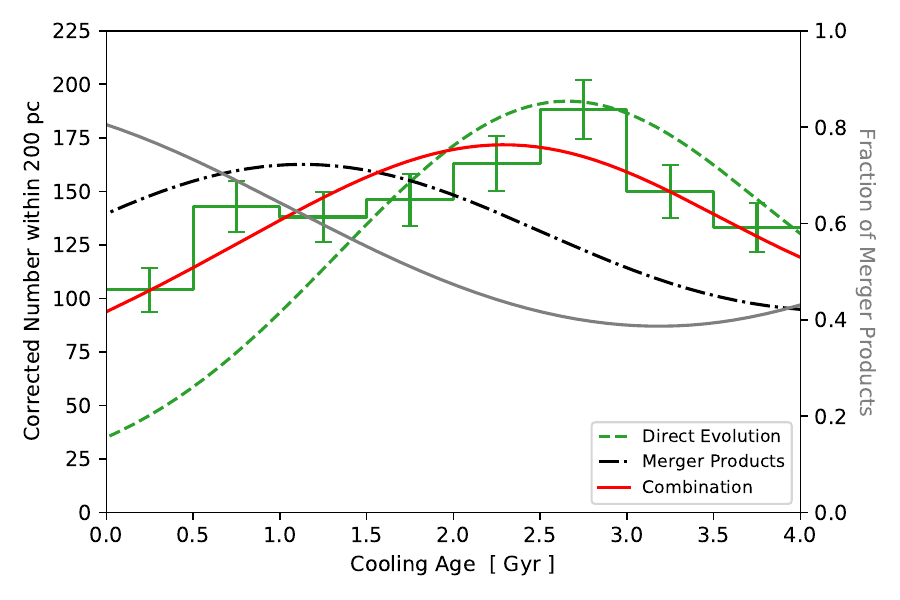}
    \caption{
    Ultramassive WD ($1.15-1.25~M_\odot$) formation mechanisms.
    In the best-fitting combination (red solid curve, $\chi_\nu^2=1.4$ with six degrees of freedom) of mergers and direct formation, 51\% of ultramassive white dwarfs that formed over the past 4~Gyr, formed through mergers. 
    Among the ultramassive WDs forming today, 80\% are the result of mergers.
    The limiting cases of only mergers (black dash-dot curve) and only single stellar evolution (dashed green curve) are shown for comparison.
    The green histogram is the empirical distribution.
    The fraction of merger products in the best-fitting combination is shown as the solid grey curve, with y-axis on the right.
    }
    \label{fig:umassive_form_mechs}
\end{figure}

In \cref{fig:umassive_form_mechs}, we show the results of fitting the different ultramassive WD formation scenarios to the empirical distribution according to the thick Montreal cooling models.
The empirical distribution is shown as the green histogram.
The solid red curve shows the cooling age distribution for the best-fitting combination of merger and direct formation products.
Also shown are the best-fitting cooling age distributions for the limiting cases in which all of the ultramassive WDs are the result of direct formation (dashed green curve) and all are the result of mergers (dash-dot black curve), which are simply proportional to the SFR and the convolution of the SFR and merger delay distribution, respectively.
Using a separate y-axis scale, \cref{fig:umassive_form_mechs} shows $f_m$ for the best-fitting combination as a solid grey line.
The scale for all of the cooling age distributions is indicated by the y-axis on the left side of the plot, while the scale for the fraction of merger products is indicated by a separate y-axis on the right side of the plot.
Comparison of the cooling age distributions visually illustrates the better fit of the optimal combination formation scenario relative to the limiting cases, as quantified by the $\chi^2_\nu$ values listed in \cref{tab:merger_combo_fits}.
For the optimal combination, we also see from \cref{fig:umassive_form_mechs} that the fraction of merger products is a function of the WD cooling age, with a maximum value for WDs that are forming today.

\section{Conclusions}

We have shown that the number density of \textit{Gaia} EDR3 WDs in the mass range $0.95-1.15~M_\odot$ gradually increases with cooling age over the first $\sim 3~\mathrm{Gyr}$ of cooling time. This pile-up of WDs extends over the entire age range associated with the Q branch and continues well beyond it, reaching a peak after the end of core crystallization. 
If the WD birthrate were constant, this apparent excess of white dwarfs would indicate an anomalous cooling delay relative to the models.
However, the discovery of a time-varying star formation rate for \textit{Gaia} DR2 main sequence stars \citep{Mor2019} suggests that the WD birthrate is likely to also vary with time, with the expectation that the distribution of cooling ages for the massive WDs that we consider closely follows this star formation rate if the WDs predominantly originate from single progenitor stars and if the models are correct.
We indeed find that WDs in the mass bins $0.95-1.05~M_\odot$ and $1.05-1.15~M_\odot$ have cumulative cooling age distributions that are statistically consistent with the expectation from the star formation rate observed for main sequence stars \citep{Mor2019}. 
We do not see statistical evidence for an anomalous cooling delay in these results.

For slightly more massive white dwarfs, in the mass bin $1.15-1.25~M_\odot$, we find that the distribution of their cooling ages is consistent with a uniform distribution for all sets of models considered over the cooling age range 1 - 2.5 Gyr, and this is not consistent with the star formation rate of \citet{Mor2019}.
This could indicate that a large fraction of these most massive WDs formed from double-WD mergers. 
The binary population synthesis results of \citet{Cheng2020} for WDs in this mass range indicate a merger delay time distribution that, when convolved with the star formation rate of the main sequence progenitors, pushes the peak density to earlier WD cooling ages and overall flattens the distribution.

We model the cooling age distribution of $1.15-1.25~M_\odot$ WDs as a linear combination of the distribution of single stellar evolution products and the distribution of double-WD merger products, and we show that this formation model provides a good fit to the empirical distribution of the photometric cooling ages determined by carbon/oxygen-core WD cooling models.
Depending on the cooling model, we find that about $69-80\%$ of the WDs forming today and $36-51\%$ of the WDs that have formed over the last 4~Gyr are the product of double-WD mergers.

We note that we find no evidence for a substantial cooling delay in the numbers of massive white dwarfs detected in \textit{Gaia} EDR3 when one takes the star formation history of the Galaxy into consideration.  For the most massive white dwarfs we find that the bulk were formed in mergers. The carbon/oxygen white dwarf models seem to fit better to the number of white dwarfs than oxygen/neon models.  This may indicate that these merger remnants are actually carbon/oxygen white dwarfs. It alternately could be that the merger delay distribution proposed by \citet{Cheng2020} is not realized in the Solar Neighbourhood, that the star formation history is not precisely that given by \citet{Mor2019}, or that the models for oxygen/neon white dwarfs require some minor revision.

\section*{Acknowledgements}

This work has been supported by the Natural Sciences and Engineering Research Council of Canada through the Discovery Grants program and Compute Canada. I.C. is a Sherman Fairchild Fellow at Caltech and thanks the Burke Institute at Caltech for supporting her research.

This work has made use of data from the European Space Agency (ESA) mission \textit{Gaia} (\url{https://www.cosmos.esa.int/gaia}), processed by the \textit{Gaia} Data Processing and Analysis Consortium (DPAC, \url{https://www.cosmos.esa.int/web/gaia/dpac/consortium}) Funding for the DPAC has been provided by national institutions, in particular the institutions participating in the \textit{Gaia} Multilateral Agreement.
We used the ``Synthetic Colors and Evolutionary Sequences of Hydrogen- and Helium-Atmosphere White Dwarfs'' website at \url{http://www.astro.umontreal.ca/~bergeron/CoolingModels/}.


\section*{Data Availability}

The data used in this paper are available through TAP Vizier and the \textit{Gaia} archive.  
We constructed the 200-pc white dwarf catalogue from \url{https://warwick.ac.uk/fac/sci/physics/research/astro/research/catalogues/gaiaedr3_wd_main.fits.gz}.
The corresponding author can also provide software to perform the analysis presented in this paper.



\bibliographystyle{mnras}
\bibliography{references}




%



\bsp	
\label{lastpage}
\end{document}